\documentclass[12pt]{spieman}  
\usepackage{amsmath,amsfonts,amssymb}
\usepackage{graphicx}
\usepackage{setspace}
\usepackage{tocloft}
\usepackage{color}
\newcommand{\ihighlight}[1]{\textcolor{black}{#1}}

\title{Dual-energy CT imaging using a single-energy CT data is feasible via deep learning}

\author[a,$\dagger$]{Wei Zhao}
\author[b,$\dagger$]{Tianling Lv}
\author[a]{Peng Gao}
\author[a]{Liyue Shen}
\author[a]{Xianjin Dai}
\author[a]{Kai Cheng}
\author[a]{Mengyu Jia}
\author[b,*]{Yang Chen}
\author[a,*]{Lei Xing}
\affil[a]{Stanford University, Department of Radiation Oncology, 875 Blake Wilbur Drive, Stanford, California, United States, 94306}
\affil[b]{Southeast University, Department of \ihighlight{Computer Science and Engineering}, \ihighlight{2 Sipailou}, Nanjing, China, 210096}
\affil[$\dagger$]{Authors contributed equally to this paper.}

\cftpagenumbersoff{figure}
\cftpagenumbersoff{table}
\begin{document}
\maketitle

\vspace{-2.5mm}
\begin{abstract}
In a standard computed tomography (CT) image, pixels having the same Hounsfield Units (HU) can correspond to different materials and it is therefore challenging to differentiate and quantify materials. Dual-energy CT (DECT) is desirable to differentiate multiple materials, but DECT scanners are not widely available as single-energy CT (SECT) scanners. Here we develop a deep learning approach to perform DECT imaging by using standard SECT data. A deep learning model to map low-energy image to high-energy image using a two-stage convolutional neural network (CNN) is developed. The model was evaluated using patients who received contrast-enhanced abdomen DECT scan with a popular DE application: virtual non-contrast (VNC) imaging and contrast quantification. The HU difference between the predicted and original high-energy CT images are 3.47, 2.95, 2.38 and 2.40 HU for ROIs on spine, aorta, liver and stomach, respectively. The HU differences between VNC images obtained from original DECT and deep learning DECT are 4.10, 3.75, 2.33 and 2.92 HU for the 4 ROIs, respectively. The aorta iodine quantification difference between iodine maps obtained from original DECT and deep learning DECT images is 0.9\%, suggesting high consistency between the predicted and the original high-energy CT images. This study demonstrates that highly accurate DECT imaging with single low-energy data is achievable by using a deep learning approach. The proposed method can significantly simplify the DECT system design, reducing the scanning dose and imaging cost.
\end{abstract}

\vspace{-1.5mm}

{\noindent \footnotesize\textbf{*}Yang Chen,  \linkable{chenyang.list@seu.edu.cn};~~
\noindent \footnotesize\textbf{*}Lei Xing,  \linkable{lei@stanford.edu}}

\begin{spacing}{1}   

\section{Introduction}
\label{sect:intro}  
In standard single-energy computed tomography (SECT) imaging, the pixel value represents effective linear attenuation coefficient and it is an averaged contribution of all materials or chemical elements in the pixel. It therefore does not give a unique description for any given material and pixels having the same CT numbers can represent materials with different elemental compositions, making the differentiation and quantification of materials extremely challenging. Dual-energy CT (DECT) scans the object using two different energy spectra and is able to take advantage of the energy dependence of the linear attenuation coefficients to yield material-specific images~\cite{kalender1986,fessler2002,carmi2005,johnson2007,la2008,matsumoto2011,maass2011,niu2014,xia2014,zhao2016,zhao2017,forghani2017,zhang2017,zhao2018}. This enable DECT to be applied in several emerging clinical applications, including virtual monoenergetic imaging, automated bone removal in CT angiography, perfused blood volume, virtual noncontrast-enhanced images, urinary stone characterization and so on, and the application list is still expanding.


In reality, vendors have used different techniques to acquire dual-energy data, including rapid kV switching~\cite{kalender1986,silva2011}, dual x-ray sources~\cite{johnson2007}, and layered detector~\cite{carmi2005,hao2013}. All these proprietary techniques have posed a significant burden on CT system hardware. Hence, DECT scanners are not widely available as SECT scanners, especially for a developing area. In addition to the increased complexity of the imaging system and cost, DECT may also increases the radiation dose to patients due to the additional CT scan. A deep learning model learns multiple levels of representations that
\newpage
\noindent
correspond to different levels of abstraction from the input images to perform prediction. It has been successfully applied into many fields, including medical imaging. In this study, instead of scanning the patient twice using different spectra, we propose to obtain DECT images from SECT by developing a deep learning strategy and show its extraordinary performance by using 22 clinical cases.

\begin{figure}
\begin{center}
\begin{tabular}{c}
\includegraphics[height=12cm]{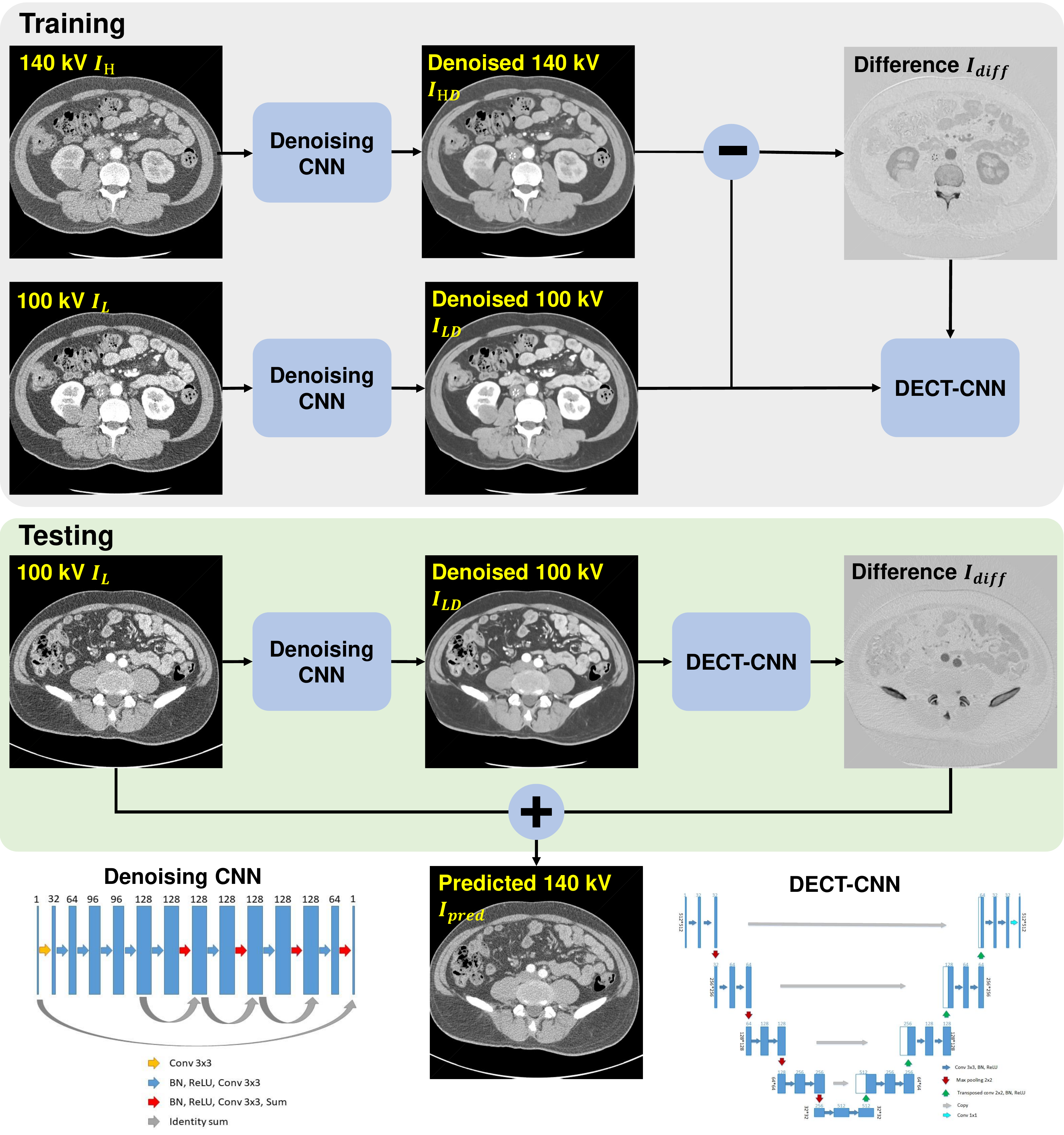}
\end{tabular}
\end{center}
\caption
{ \label{fig:ct}
Overall flowchart of the proposed deep learning-based DECT imaging method.}  
\vspace{-1.5mm}
\end{figure}

\section{Methods}

Suppose $I_H, I_L$ are the low-energy and the corresponding high-energy CT images, respectively. We aim to train a deep learning model to transform $I_L$ to $I_H$ using paired DECT images. The end point of the deep learning approach was a model capable of providing the high-energy CT image $I_H$ for a given input low-energy CT image $I_L$. To this end, we first used a dedicated ene-to-end convolutional neural network (CNN) model~\cite{yang2017} which is \ihighlight{a} fully convolutional network derived from ResNet to yield noise significantly reduced high- and low-energy CT images $I_{HD}$ and $I_{LD}$. Instead of directly mapping the high-energy CT image from the low-energy CT image, we then trained an independent mapping CNN, which was referred to as DECT-CNN, to learn the difference $I_{diff}$ between $I_{HD}$ and $I_{LD}$ for the given input $I_{LD}$, and the predicted high-energy CT image $I_{pred}$ is calculated as the summation of the origianl low-energy CT image $I_L$ and the predicted difference image $I_{diff}$ during inference procedure. The workflow of proposed deep learning-based DECT imaging method is shown in \ihighlight{Fig.~\ref{fig:ct}}.

Due to the complexity of predicting $I_{diff}$ from $I_{LD}$, a large reception field is needed for the DECT-CNN model. We therefore use the U-Net structure for the DECT-CNN to take advantage of its extremely large reception field and the ability to keep the high-resolution context information. Different from the original U-Net which uses softmax and cross entropy loss, we have updated the DECT-CNN model by using a mean-squared-error loss.

\begin{figure}
\begin{center}
\begin{tabular}{c}
\includegraphics[height=3.5cm]{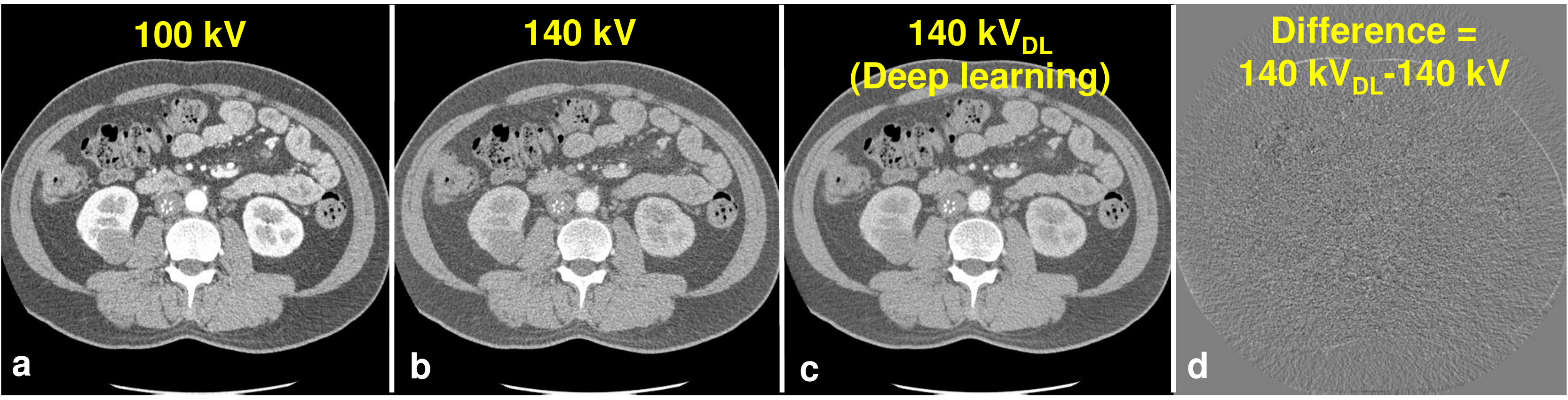}
\end{tabular}
\end{center}
\caption
{ \label{fig:vnc}
Original (a) low- and (b) high-energy DECT images and (c) predicted high-energy CT images, and (d) difference image between the predicted and original 140 kV images. All images are displayed in (C=0HU and W=500HU).}  
\end{figure}

Based on the original low-energy CT image $I_L$ and the predicted high-energy CT image $I_{pred}$, we study the performance of the deep learning-based DECT using a popular DE application: virtual non-contrast (VNC) imaging and contrast quantification. A statistically optimal image-domain material decomposition algorithm is employed to obtain the VNC images and the iodine maps~\cite{faby2015}. To assess the accuracy of the approach, we retrospectively studied DECT images from 22 patients who received contrast-enhanced abdomen CT scan. The proposed model was trained on 16 randomly selected image volumes which include 4949 image slices.  The remaining 6 volumes were equally divided into a validation group and a testing group each of which includes 3 volumes. Quantitative comparisons between the predicted high-energy CT images and the original high-energy CT images were performed using HU accuracy in 60 region-of-interests (ROIs) on different types of tissues. VNC images and iodine maps quantification obtained from the original DECT images and the deep learning-based DECT images were compared and quantitatively evaluated using HU value and noise level.

\vspace{-0.2mm}
\section{Results}

\begin{figure}
\begin{center}
\begin{tabular}{c}
\includegraphics[height=11.5cm]{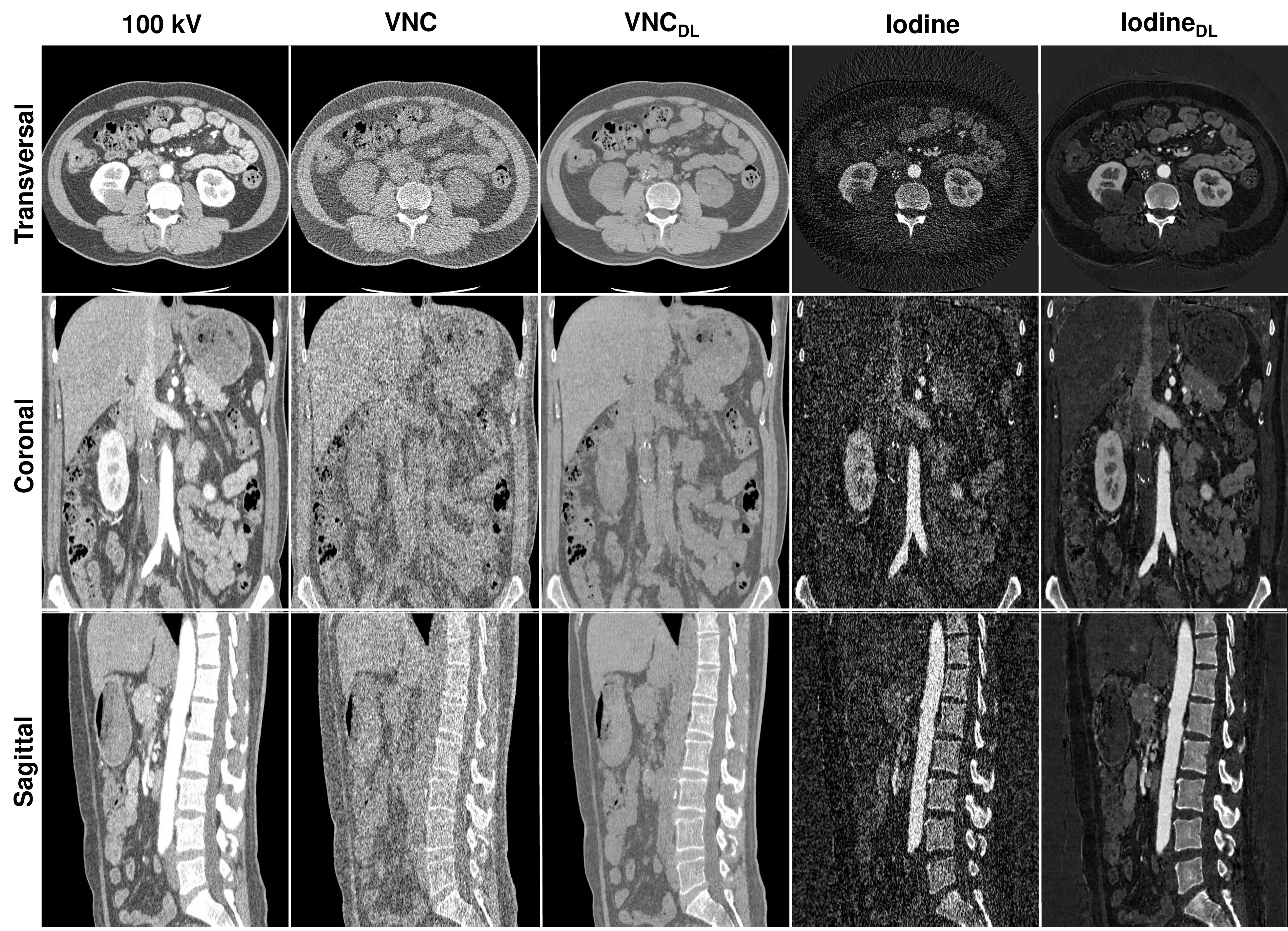}
\end{tabular}
\end{center}
\caption
{ \label{fig:example}
Illustration of the contrast-enhanced 100 kV CT images (C=0HU and W=500HU) and VNC images (C=0HU and W=500HU) and iodine maps (C=0.6 and W=1.2) obtained using original DECT images and deep learning (DL)-based DECT images in transversal, coronal and sagittal views.}  
\end{figure}

We found that the mapping CNN yields inferior high-energy CT images with original noisy low-energy CT images as input, indicating the CNN wastes its ability of expression on mapping the image difference between high- and low-energy levels with the presence of noise. With noise significantly reduced DECT images, the CNN can appreciate the ingenuous CT image difference at different energy levels and ultimately yield superior high-energy CT image. \ihighlight{An example image slice is shown in Fig.~\ref{fig:vnc}.} The HU difference between the predicted and original high-energy CT images are 3.47 HU, 2.95 HU, 2.38 HU and 2.40 HU for ROIs on spine, aorta, liver and stomach, respectively. \ihighlight{Fig.~\ref{fig:example}} shows the three-dimensional VNC images and iodine maps obtained from original 100 kV/140 kV DECT images and deep learning-based DECT images. As can be seen, the deep learning-based DECT approach provides high quality VNC and iodine maps. Since material decomposition uses matrix inversion and yields amplified image noise, the noise levels in VNC images and iodine maps obtained from the original DECT images are much higher than the 100 kV images. Due to the noise correlation between the predicted high-energy CT images and the original low-energy CT images, deep learning-based DECT imaging provides noise significantly reduced VNC images and iodine maps. The HU differences between VNC images obtained from original DECT and deep learning DECT are 4.10 HU, 3.75 HU, 2.33 HU and 2.92 HU for ROIs on spine, aorta, liver and stomach, respectively. The aorta iodine quantification difference between iodine maps obtained from original DECT and deep learning DECT images is 0.9\%, suggesting high consistency between the predicted and the original high-energy CT images.

\section{Discussion and Conclusion}

Since we have used an image domain material decomposition method, artifacts in the CT images, such as beam hardening artifacts, scatter artifacts, may reduce the accuracy of the material-specific images. This can impact material decomposition using both the original DECT images and the deep learning DECT images. Hence, in the cases where artifacts significantly reduce the HU accuracy of CT images, beam hardening correction~\cite{zhao2012,zhao2019} and scatter correction~\cite{Niu2011,zhao2014scatter,zhao2015scatter,zhao2016scatter,shi2019} methods can be employed to reduce the CT artifacts and further to enhance the material decomposition accuracy.

This study demonstrates that highly accurate DECT imaging with single low-energy data is achievable by using a deep learning approach. The proposed algorithm shows superior and reliable performance on the clinical datasets, and provides clinically valuable high quality VNC and iodine maps. Compared to the current standard DECT techniques, the proposed method can significantly simplify the DECT system design, reduce the scanning dose by using only a single kV data acquisition, and reduce the noise level of material decomposition by taking advantage of the noise correlation of the deep learning derived DECT images. The strategy reduces the DECT imaging cost and may find widespread clinical applications, including cardiac imaging, angiography, perfusion imaging and urinary stone characterization and so on.





\bibliographystyle{spiejour}   





\end{spacing}
\end{document}